\documentclass{ab}

\usepackage{graphicx}
\usepackage{natbib}
\newcommand{\km}{\mbox{km}\,\mbox{s}^{-1}}

\begin{document}

\title{Reduction of CCD observations made with a scanning Fabry--Perot interferometer. III.~Wavelength scale refinement}

\author{A.~V. Moiseev}

\institute{Special Astrophysical Observatory, Russian Academy of Sciences, Nizhnij Arkhyz, 369167, Russia}
 \titlerunning{Reduction of observations with a scanning F-P Interferometer}

\authorrunning{Moiseev}

\date{August 10,  2015/Revised: Neptember 7, 2015}
\offprints{A. Moiseev  \email{moisav@sao.ru} }

\abstract{
We describe the recent modifications to the data reduction
technique for observations acquired with the scanning Fabry--Perot
interferometer (FPI) mounted on the 6-m telescope of the Special
Astrophysical Observatory that allow the wavelength scale to be
correctly computed in the  case of large mutual offsets of studied
objects in interferograms. Also the parameters of the scanning FPIs used in the SCORPIO-2 multimode focal reducer are considered.
}

\maketitle

\section{Introduction}
\label{intro}

Integral-field (3D) spectroscopy is now used extensively both in
optical and IR observations on telescopes of various sizes.
The most popular instruments employed are integral-field
spectrographs and units (IFU) equipped with microlens arrays,
fiber bundles, or slicers. However, the field of view provided by
this technique barely reaches about one arcminute even in the best
cases MUSE~\citep{MUSE} on the \mbox{8-m} VLT telescope
or PPAK/PMAS~\citep{PPAK} on the 3.6-m Calar Alto
telescope. This limitation is due to the need to simultaneously
record many spectra with sufficient spatial sampling. At the same
time, the spectrometers based on scanning Fabry-Perot interferometer (FPI) mounted on 2--10-m telescopes prove to be
unchallenged when it comes to a combination of such parameters as
a large (several arcminutes) field of view with spatial sampling
similar to that of direct images and a relatively high
(\mbox{$R=\lambda/\delta\lambda>10\,000$}) spectral
resolution~\citep{Boulesteix2002}. The downside of these
advantages is the narrow spectral interval (usually less than
50~\AA) and the need to successively record individual
interferogram frames (scanning or  multiplex spectrometer).
Account should also be taken of atmospheric transparency and
seeing variations over long scanning cycles, which can often
continue for several hours. Two approaches are employed to address
these problems: the use of photon counting detectors, which allow
multiple repeated scannings with short exposures, or the ``slow''
scanning with a high quantum efficiency CCD with finite readout
time and noise levels. An example of the first approach is the
GH$\alpha$FaS instrument mounted on the 4.2-m WHT
telescope~\citep{Hernandez2008} and equipped with a GaAs
photo-cathode and a microchannel plate based amplifier at the
input of a commercial CCD. The second approach, the one involving
a classic CCD, allows the same detector to be used for other types
of observations such as direct imaging and long-slit spectroscopy,
i.e., incorporating the FPI into a multimode instrument---the
focal reducer. It is important to develop an observing and data
reduction technique involving proper correction of atmospheric
modulations in the case of slow scanning with an FPI. This
approached is implemented on the 6-m telescope of the Special
Astrophysical Observatory of the Russian Academy of Sciences
(SAO~RAS), where the FPI has been used as one of the operation
modes of the SCORPIO focal reducer since
2000~\citep{AfanasievMoiseev2005}.

For a detailed description of the data reduction technique used in
observations with the FPI mounted on SCORPIO, see our previous
papers of this series~\citep{Moiseev2002ifp,
MoiseevEgorov2008}. In this paper we describe further
progress in the application of this technique on the 6-m telescope
for studies of ionized gas kinematics in Galactic and
extragalactic objects. For instance, in 2010  first light was
achieved on the SCORPIO-2 second-generation multimode focal
reducer~\citep{AfanasievMoiseev2011}, which includes the
FPI mode. Since 2013 all such observations on the \mbox{6-m}
telescope have been performed  with SCORPIO-2 exclusively. Two new
scanning piezoelectric interferometers were acquired, and the
spectral resolution has been almost doubled.

\begin{figure} 
\includegraphics[scale=0.5]{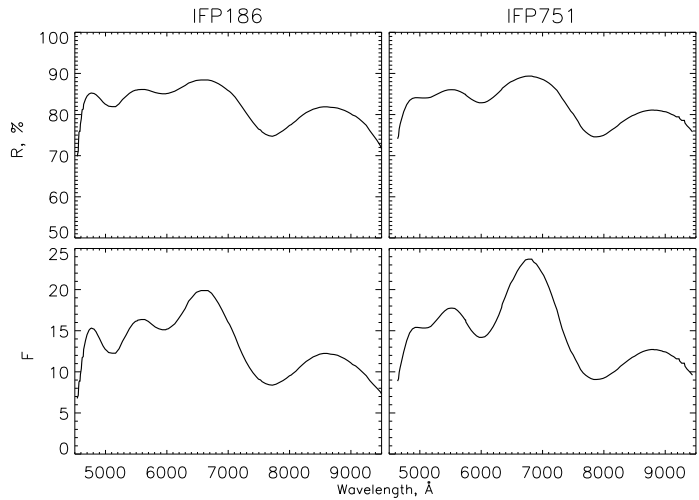}
\caption{Wavelength dependence of the reflection coefficient $R$
of interferometer plates (top) and the corresponding computed
finesse $F$ (bottom). The left and right panels show the data for
the IFP186 and IFP751 interferometers respectively.}
\label{fig1}
\end{figure}

In the first part of this paper (Section~\ref{sec2}), we describe the parameters of the
FPIs used on SCORPIO-2. The emergence of new hardware and the
diversification of research tasks required a modification of the
data reduction algorithm. Section~\ref{sec3}  discusses
the factors that cause wavelength scale errors in the case of
frame-to-frame offsets of studied objects  in interferograms.
Section~\ref{sec4} describes a method for correcting
this effect: such a correction is important for precise ionized
gas velocity dispersion measurements. The concluding section
discusses  possible applications of the method developed.

\section{New Fabry--Perot interferometers at SAO~RAS}

\label{sec2}

The principal parameters of FPI observations are practically
identical both for SCORPIO and SCORPIO-2: the same field of view ($6\farcm1 \times 6\farcm1$) with a $0\farcs36$/pixel scale (in the
case of  $2\times 2$ binned CCD readout). The differences include
a somewhat higher sensitivity of the E2V-4240 CCD in the H$\alpha$
region (95\% compared to  78\% for the \mbox {EEV~42-40} CCD used
in the old spectrograph). SAO~RAS also acquired two new
\mbox{ET-50} scanning piezoelectric interferometers with serial
numbers FS-1064 and \mbox{FS-1081} (manufactured in 2009 and 2012
respectively) from  IC Optical Systems,~Ltd.\footnote{\tt
http://www.icopticalsystems.com/}. In the
SCORPIO-2 nomenclature these interferometers are referred to as
IFP186 and IFP751 respectively. The table~\ref{tab_1} lists the main
parameters of the devices: $n$, the order of interference at the
given wavelengths; $\Delta\lambda$, the free spectral range
between two adjacent orders; $\delta\lambda$, the spectral
resolution ($\rm FWHM$ of the line profile of the calibration
lamp); $F$, the finesse; $n_z$, the number of images (channels) in
the spectral interval. The last parameter is set during
observations so as to ensure the minimum allowable sampling, i.e.,
\mbox {$n_z\geq2F$}.

\begin{table}[b]
\caption{\centerline{Parameters of the scanning FPIs used with SCORPIO-2}} 
\label{tab_1}
\medskip
\begin{tabular}{l|c|c|c|c}
\hline
Parameter  &  \multicolumn{2}{c|}{IFP186}         & IFP751        & IFP501          \\
           & $\lambda\,6563$ & $\lambda\,5007$& $\lambda\,6563$  & $\lambda\,6563$ \\
\hline
$n$                 & 188            & 246         & 751              &  501         \\
$\Delta\lambda$, \AA& 34.9           & 20.3        & 8.7              &  13.1        \\
 $\Delta\lambda$, $\km$    & 1696         & 1216      & 399            &  598   \\
$\delta\lambda$, \AA& 1.7            & 2.0         & 0.44             &  0.80        \\
 $\delta\lambda$, $\km$   & 78           & 120       & 20             &  36    \\
$F$                 & 21             & 10          & 20               &  16          \\
$n_z$               & 40             & 30          & 40               &  36          \\
 \hline
\end{tabular}
\end{table}

\begin{figure*}
\includegraphics[scale=0.9]{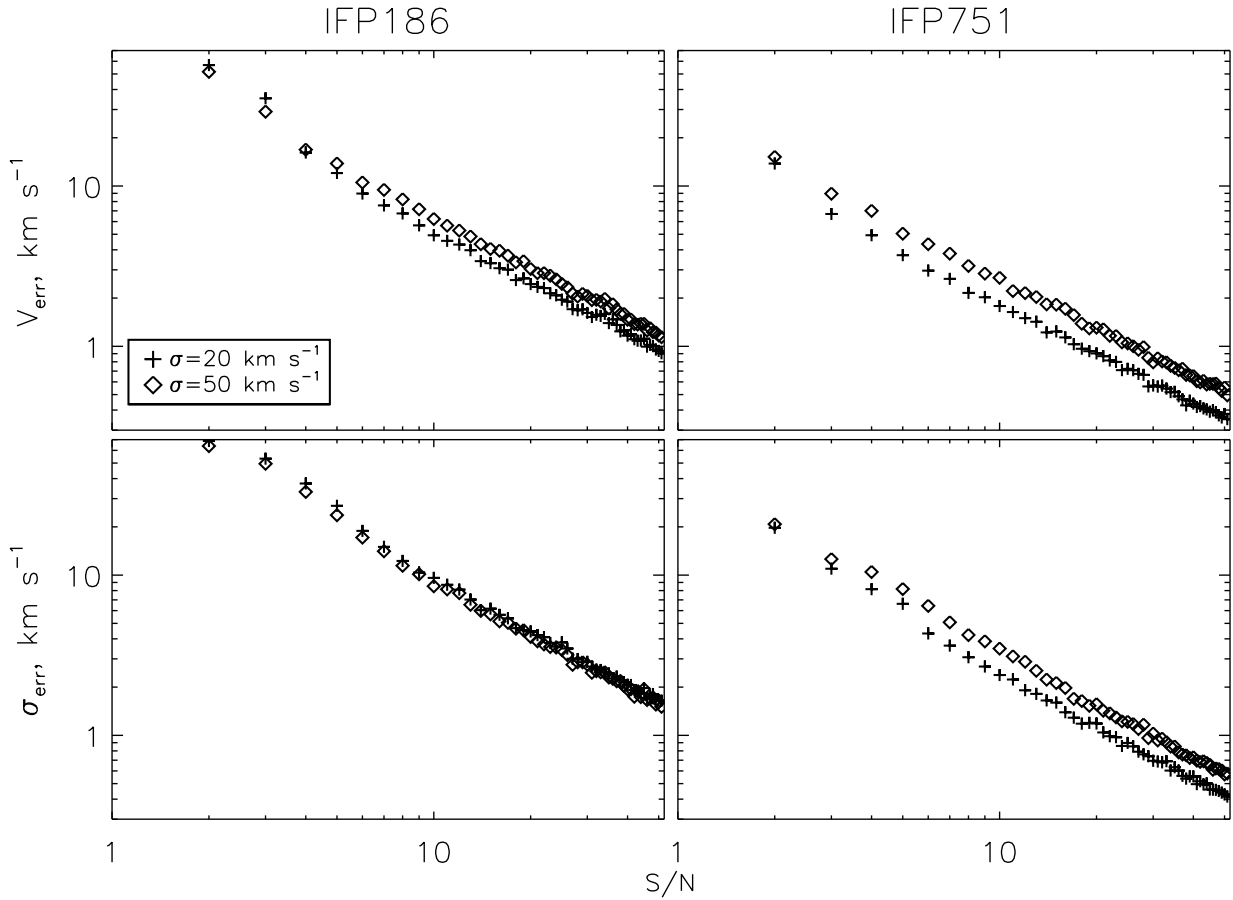}
\caption{Simulation of the measurement errors of kinematic
parameters (at the  $1\sigma$ level) in the case of the
Voigt-profile approximation of the  H$\alpha$ line: the dependence
of the error of the measured radial-velocity (top) and velocity
dispersion (bottom) on the signal-to-noise ratio. The
corresponding data are provided both for the IFP186 (left) and
IFP751 (right) interferometers. The crosses and diamond signs show
the results of calculations for velocity dispersions of 20 and
50~$\km$ respectively.} \label{fig2}
\end{figure*}

These parameters were measured directly with \mbox{SCORPIO-2} from
the spectrum of the He-Ne-Ar  calibration lamp with
individual lines separated using narrow-band filters. The
$\delta\lambda$ parameter in the vicinity of H$\alpha$ was
estimated from the Ne\,I~$\lambda$\,6598.95 line, and that in the
vicinity of the [O\,III] line---by simultaneously fitting the
profiles of five Ne\,I\,+\,Ar\,II lines in the  \mbox
{5145--5152}~\AA{} wavelength interval. The manufacturer provided
the plot of the measurements of the reflection coefficient $R$ of
interferometer plates as a function of wavelength
(Fig.~\ref{fig1}, top panels). Also shown in the same figure are the
corresponding finesse characterizing the spectral resolution at
different wavelengths as determined by formula
\begin{equation}
F(\lambda)=k\,F_0(\lambda).
\end{equation}
Here  $F_0(\lambda)$ is the finesse of ``ideal'' interferometer
(with no defects and absolutely plane plates) computed by the
following formula, known from FPI theory:
\begin{equation}
F_0(\lambda)\approx\displaystyle{\frac{\pi\sqrt{R(\lambda)}}{1-R(\lambda)}};
\end{equation}
and the normalization factor $k$ chosen so as to ensure the best
agreement between the resulting plot and both our measurements
(see Table) and manufacturer's data ($F=18$ at  633~nm for both
interferometers): $k=0.78$ and $k=0.85$ for IFP186 and IFP751
respectively. Figure~\ref{fig1} shows that observations
with a sufficiently high finesse  ($F>10$) can be performed not
only near the H$\beta$--[O\,III]~$\lambda\,5007$ and
H$\alpha$--[S\,II]~$\lambda\lambda\,6717$, $6731$ lines but also
in the vicinity of the  Ca\,II $\lambda\lambda\,8498$, $8542$,
$8662$ triplet. This possibility was incorporated intentionally so
that the instrument could be used to study stellar population
kinematics by analyzing absorption lines.

\begin{figure}
\includegraphics[scale=0.7]{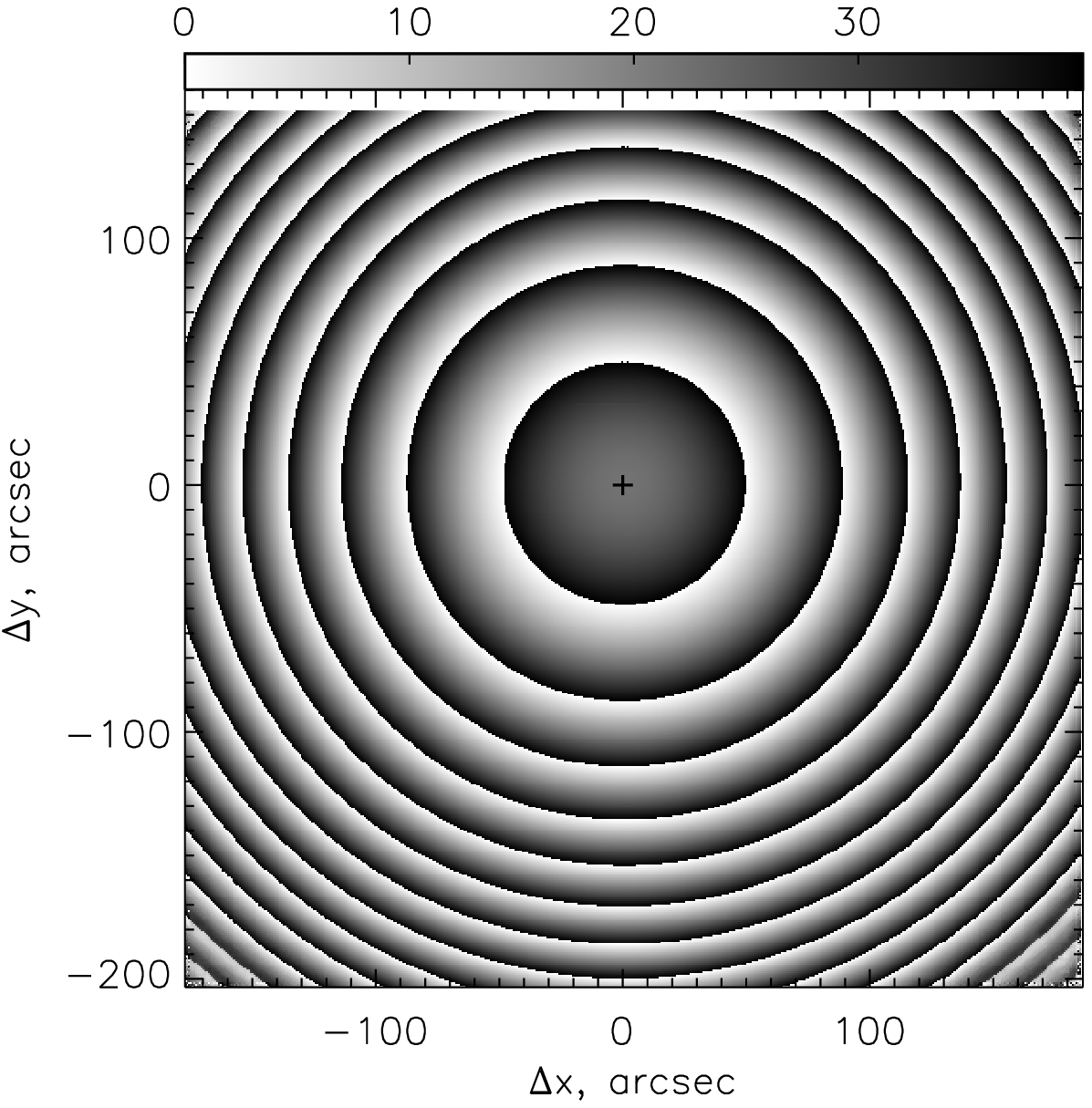}
\includegraphics[scale=0.7]{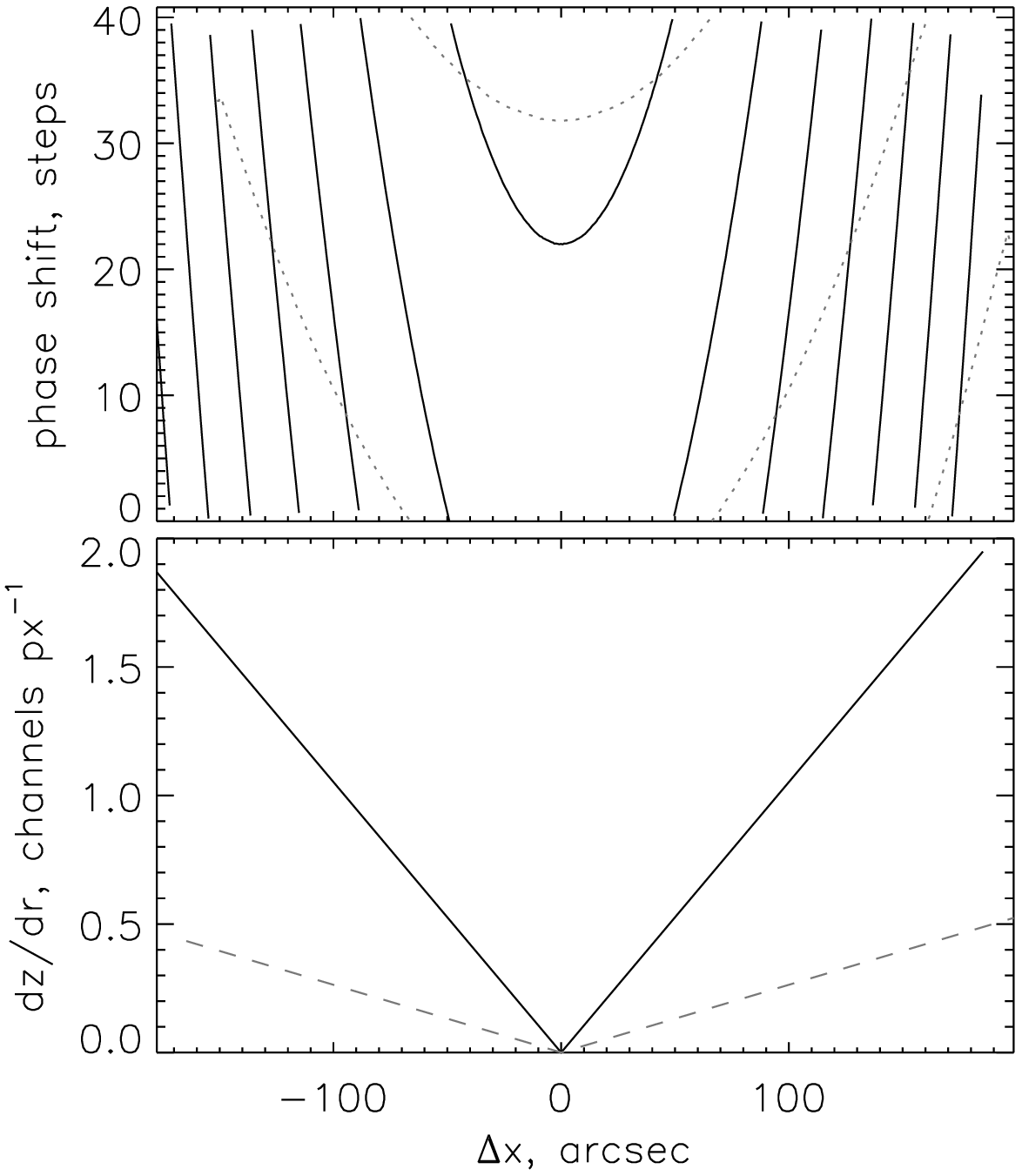}
 \caption{Top: a phase map based on the measurements of the
Ne\,I~$\lambda\,6598.95$ line in the spectrum of the calibration
lamp (in units of spectral channels for IFP751). Middle:
horizontal sections of phase maps passing through the center of
the fringe pattern for IFP751 (the solid line) and IFP186 (the
dotted line). Bottom: the corresponding variations of angular
dispersion along the radius (the same designations are used). }
\label{fig3}
\end{figure}

We provisionally computed the order of interference $n$ for a
given wavelength based on the gap between the interferometer
plates as measured by the manufacturer (with an accuracy of about
3~$\mu$m). We then refined this quantity by measuring the
positions of neighbouring  (with a wavelength difference of
\mbox{2--3\,$\Delta\lambda$}) lines in the spectrum of the
calibration lamp  so that the wavelength difference measured after
scanning would be as close as possible to value from the literature. For
measurements with the available narrow-band filters and the
He-Ne-Ar lamp, we selected line pairs in the 5080--5220~\AA{}
interval. The accuracy of such gap measurements is equal to \mbox
{$n\pm1$}  (about $\lambda/2$), which is quite sufficient for
further wavelength scale calibration in observations involving a
sole calibrating line. Notice that during the first observations
performed with IFP186, the order of interference was estimated as
\mbox {$n(\lambda\,6563)=186$}. It was later refined to be 188,
but the nomenclature was left unchanged.

For comparison, we also list in the table~\ref{tab_1} the parameters of the
old interferometer IFP501. It is practically hardly ever used now,
because IFP751 provides almost twice better spectral resolution
for a similar~$\Delta\lambda$.  IFP186 is more convenient for
observations of faint objects with a wide range of radial
velocities. In particular, it is used for observations of
active galactic nuclei in the  [O\,III]~$\lambda\,5007$ line.

Figure~\ref{fig2} shows the estimated measurement
accuracy for main emission-line kinematic parameters, the radial
velocity $v$ and velocity dispersion $\sigma$, plotted as
functions of the signal-to-noise ratio $S/N$.  Using  a technique
described in~\citet{MoiseevEgorov2008}  we modeled the instrumental
profile of the FPI by the Lorentz function with the $\rm FWHM$
adopted from the table~\ref{tab_1} for the case of H$\alpha$
observations. The wavelength scale parameters ($\Delta\lambda$ and
$n_z$) were also adopted from the table~\ref{tab_1}. We smoothed the profile
by a Gaussian with a given dispersion  $\sigma$, then superimposed
noise and estimated $v$ and $\sigma$ by approximating the line by
the Voigt profile. We performed 1000 independent trials for each
fixed $S/N$. The figure shows the measurements made for velocity
dispersions \mbox{$\sigma=20$} and 50~$\km$, which corresponds to
the typical range of this parameter in  H\,II regions in galaxy
disks \citep[see, e.g.][]{Moiseev2015}. It is evident
from the figure that for the FPIs employed, the measurement
accuracy for both kinematic parameters is better than 2--4~$\km$
for $S/N>20$.

SCORPIO-2 observations with the FPI are performed with ${\rm
FWHM}=13$--$35$~\AA-wide narrow-band filters cutting a narrow
spectral interval around the line considered. Observations in the
[O\,III]~$\lambda\,5007$ and H$\alpha$ lines can be performed for
objects with radial velocities ranging from  $-300$ to
13\,000~$\km$. The filter set is systematically upgraded with the
actual list available on the web page of the
instrument.\footnote{\tt
http://www.sao.ru/hq/lsfvo/devices/scorpio-2/}

\section{Wavelength scale errors}

\label{sec3}

The estimates shown in Fig.~\ref{fig2} are the lower
limits because they refer to the ideal case of uniform noise,
which fails in the presence of bright telluric emission lines.
Subtraction of night-sky lines may produce artifacts in reduced
spectra. It is also important how accurately the instrumental
profile of the interferometer is reproduced during scanning. For a
discussion of how the choice of a subtraction algorithm affects
the resulting spectrum see \citet{Moiseev2002ifp}. Below
we analyze the factors that cause errors in the wavelength scale
of observed spectra, which also depend on the adopted reduction
technique.

Hereafter by  ``channels'' we mean the individual two-dimensional
interferograms that make up the observed data cube $I(x,y,z)$,
where $x,y$ are the CCD coordinates, and $z$ is the channel number
proportional to the gap between the FPI plates. The reduction
sequence can be subdivided into the following main steps
\citep[see][for details]{Gordon2000,Moiseev2002ifp}.
\begin{list}{}{
\setlength\leftmargin{2mm} \setlength\topsep{2mm}
\setlength\parsep{0mm} \setlength\itemsep{2mm} }
 \item (1) Assembling the data cube from observed interferograms $I(x,y,z)$,
which includes standard CCD frame reduction procedures: bias and
dark frame subtraction and flat-fielding.
 \item (2) Channel-by-channel airglow line subtraction.
  \item (3) Channel correction based on the photometry of field stars:
  compensation of mutual frame offsets, atmospheric transparency and seeing variations.
  \item (4) Conversion of the spectra to the wavelength scale, i.e.,
transition to the $I(x,y,\lambda)$ data cube. If a photon counter
is used, the sky background subtraction can be performed after
this stage~\citep{Daigle2006}.
\end{list}

Conversion to the wavelength scale is conveniently performed based
on the ``phase offset'' map $p(x,y)$, which shows the positions of
the line center in the cube made up of the interferograms of the
selected line in the spectrum of the calibrating lamp
($\lambda_{\rm calib}$). Figure~\ref{fig3} shows
examples of such maps for the high- and low-resolution SCORPIO-2
interferometers. Phase offset $p$ varies almost quadratically with
the distance  $r$ from the center of the rings. Sharp
discontinuities correspond to the change of the interference
order. If the phase map is based on the measurements of a
calibration lamp line that is close (in terms of~$\Delta\lambda$)
to the spectral line of the observed object, then in practice the
transition $I(x,y,z)\rightarrow I(x,y,\lambda)$ reduces, up to a
certain constant, to shifting each spectrum in the cube by
$-p(x,y)$.

Problems with accurate wavelength scale calibration arise in the
case of substantial mutual image offsets between individual
channels. These offsets are primarily due to instrumental
flexures, because the position of the telescope and the
parallactic angle may change appreciably over several hours of
observations while all channels are scanned.

\begin{figure*}
\centerline{
\includegraphics[scale=0.7]{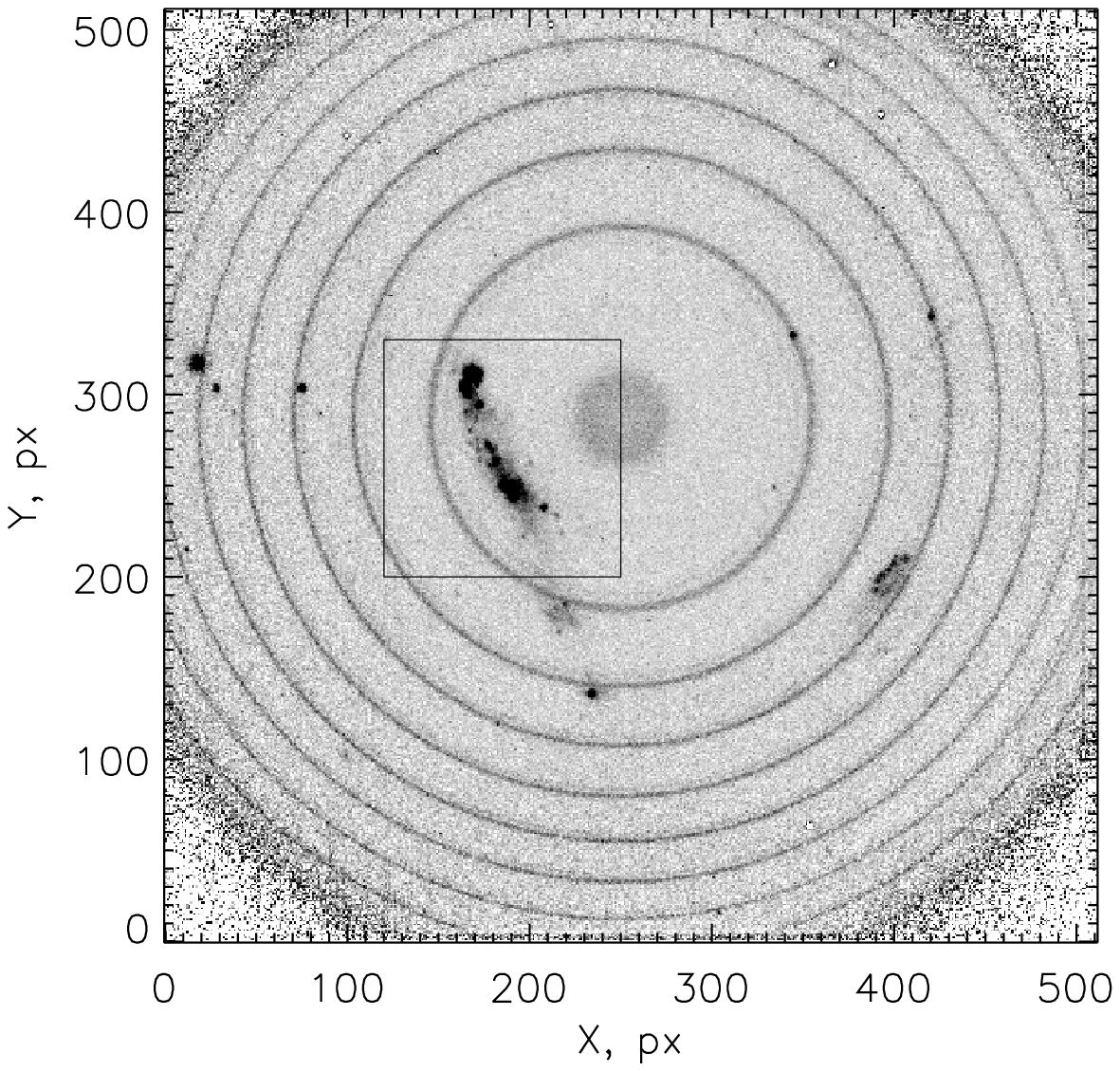}
\includegraphics[scale=0.7]{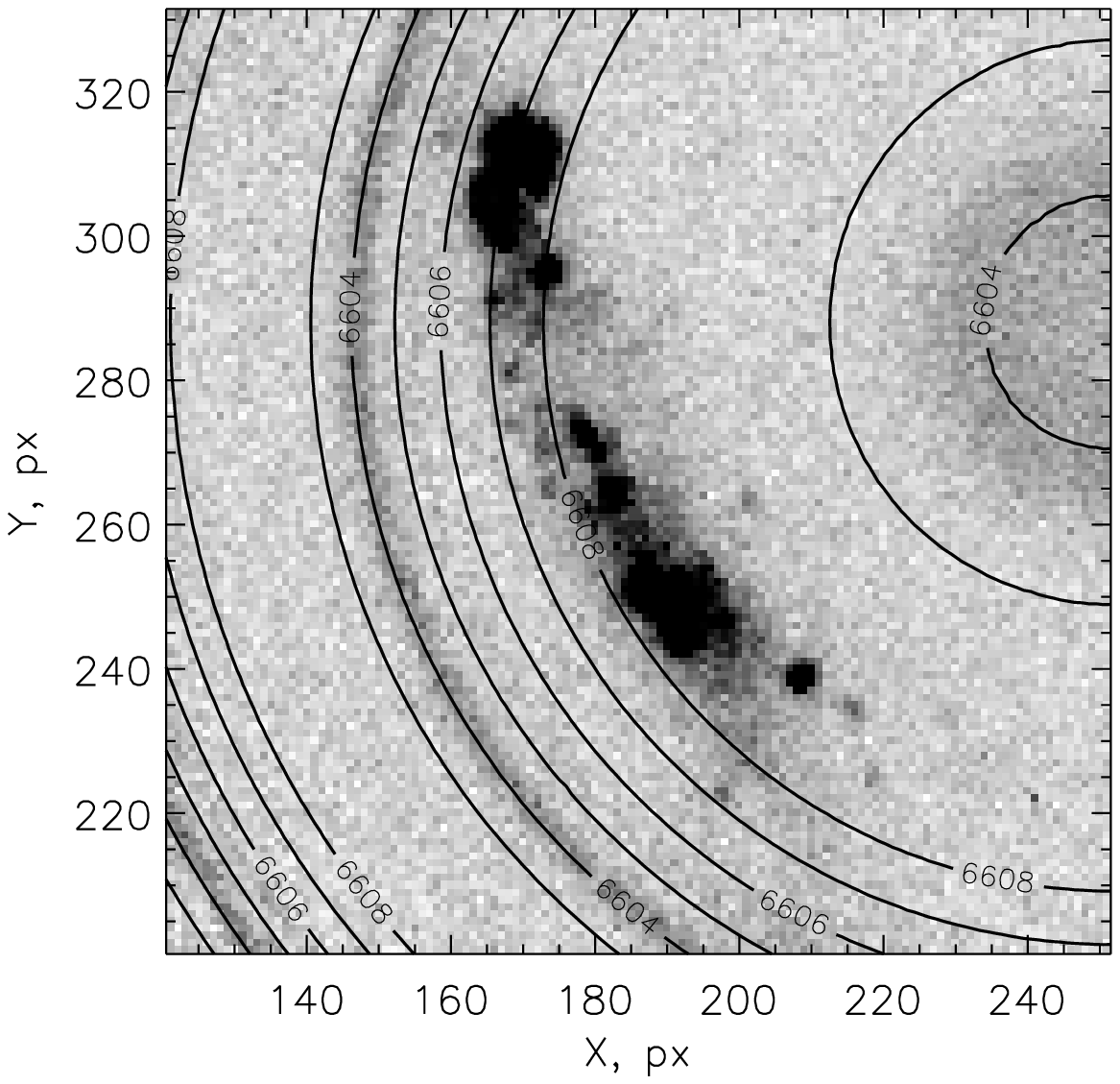} }
\caption{H$\alpha$ observations of the galaxy UGC\,260 with IFP751
mounted on SCORPIO-2. Left: an example of a frame with both the
emission from the galaxy and the rings from the $\lambda$\,6604
airglow line are immediately apparent. The square area is shown in
the right panel, the contours indicate the wavelength scale with a
contour step of 1~\AA.} \label{fig4}
\end{figure*}

Each channel contains a mixture of spatial and spectral
information, because $\lambda$ decreases with $r$
(Fig.~\ref{fig4}). Angular dispersion increases linearly
with radius and is independent of the interference order. It can
be easily shown that
\begin{equation}
\displaystyle{\frac{d\lambda}{d\vartheta}}=- \lambda
\tan\vartheta,
\end{equation}
where $\vartheta$ is the distance from the center in angular
units. The dispersion expressed in units of scanning steps is
proportional to the order of interference:
\begin{equation}
\displaystyle{\frac{dz}{d\vartheta}}=\displaystyle{\frac{n_z}{\Delta\lambda}\frac{d\lambda}{d\vartheta}}=-nn_z\tan\vartheta,
\end{equation}
as illustrated by the bottom plot in Fig.~\ref{fig3}. In
observations with IFP751, mutual channel offsets of about one
pixel ($0\farcs7$) prove to be critical even for objects in the
central part of the frame ($r\le100''$). In this case the formal
application of channel-by-channel correction procedures  (stage~3)
shift the data point in the spectrum by one channel.

\begin{figure}
 \includegraphics[scale=0.7]{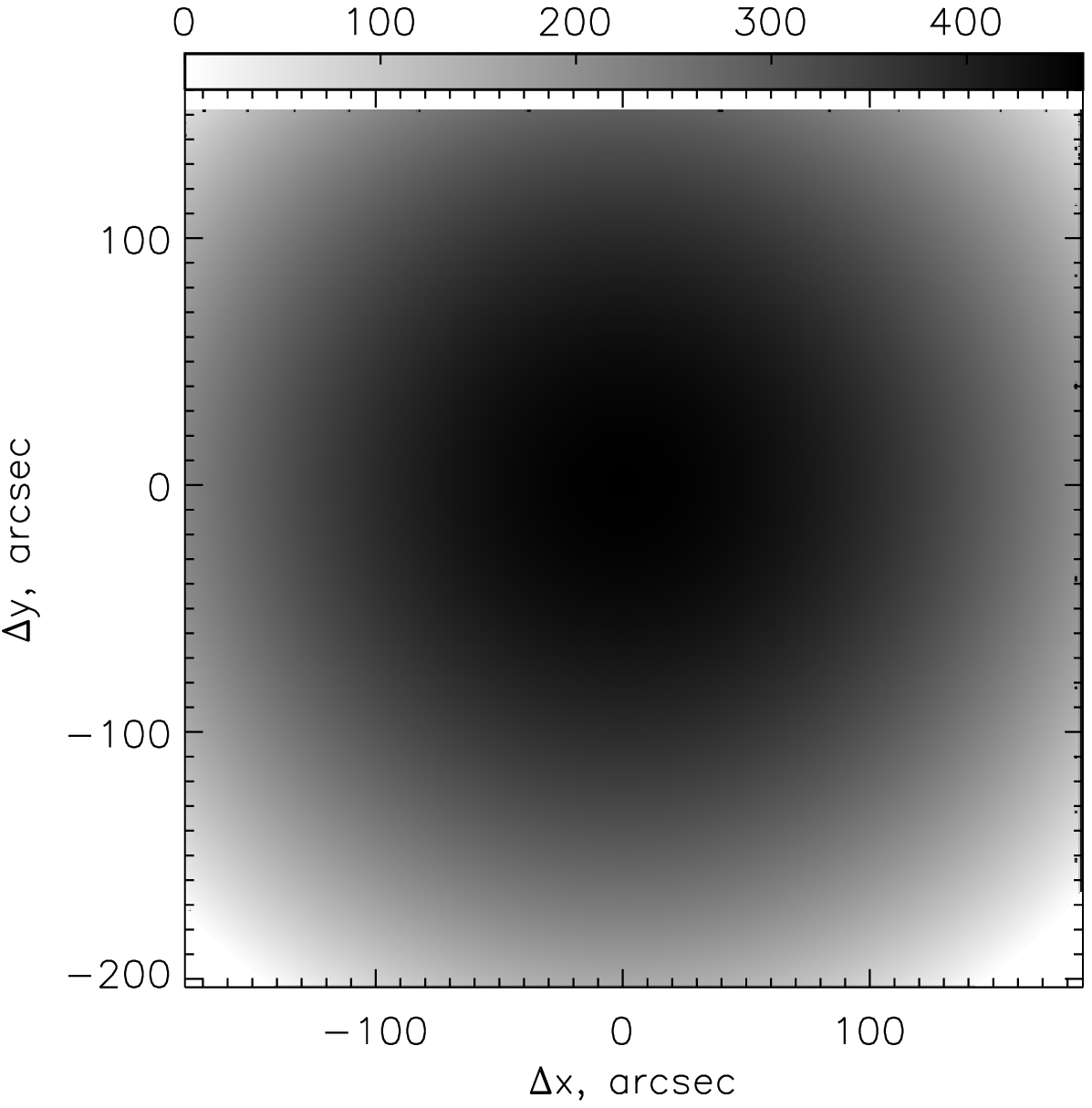}
\vbox{
\hspace{2.5mm}\includegraphics[scale=0.7]{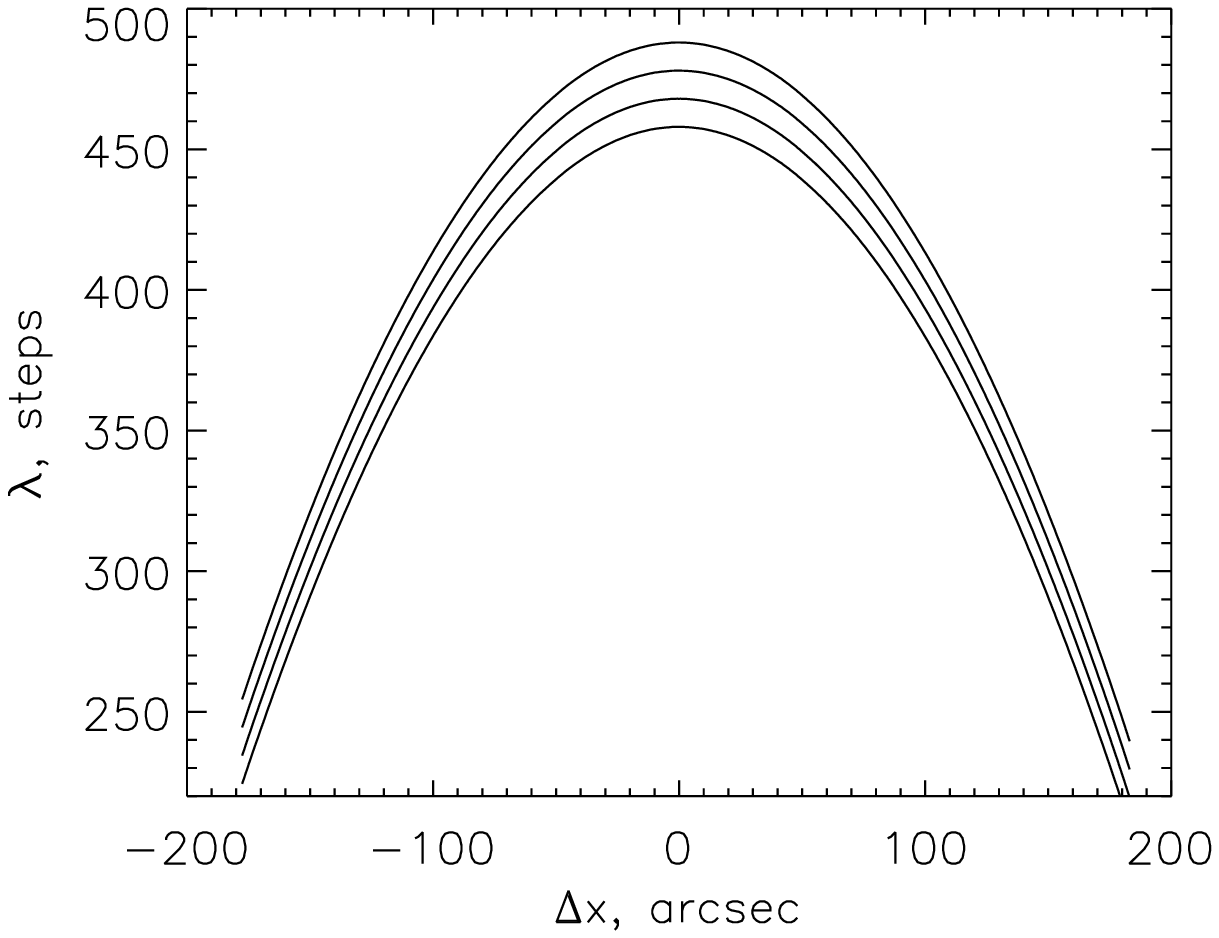}}
\caption{The \mbox{$\Lambda$-cube} constructed from the phase map
shown in Fig.~\ref{fig3}. Top: the first-channel image
with the brightness scale in scanning steps. Bottom: examples of
horizontal sections of the  \mbox{$\Lambda$-cube} passing through
the center of the rings. From bottom to top: channels 1, 11, 21,
and 31. } \label{fig5}
\end{figure}

Such offsets usually do not exceed  0.5~pixel over two-to-three
scanning hours. However,  they exceeded 1--2~pixels in some cases
(play in the carriage of the SCORPIO dispenser). The adopted
observing sequence with the odd channels recorded first and then
followed by even channels partially removes the effect due to
monotonic variation of offsets with time. However, these offsets
may nevertheless affect at least the width of the resulting line
profile.

\section{$\Lambda$-cube based correction}

\label{sec4}

We proposed the following modification of the data reduction
algorithm to avoid the above problems. Instead of the
two-dimensional phase map our new procedure uses the
$\Lambda(x,y,z)$ cube (hereafter referred to as the
\mbox{$\Lambda$-cube}) that stores the wavelength values for each
data point in $I(x,y,z)$:
\begin{equation}
\Lambda(x,y,z)=[z-p(x,y)]\displaystyle{\frac{\Delta\lambda}{n_z}}+\lambda_{\rm
calib}. \label{eqL}
\end{equation}
Figure~\ref{fig5} shows an example of such a cube. In
practice, wavelengths can be conveniently expressed in fractions
of the scanning step (the bracketed expression in
formula~(\ref{eqL})). The $z=n_z$ quantity is added to
each successive order, and hence the resulting
\mbox{$\Lambda$-cube} contains no discontinuities due to the
change of the interference order.

If  $\delta x$, $\delta y$ channel offsets have to be applied to
the object cube $I(x,y,z)$ in the process of primary reduction,
then the same offsets should also be applied to the corresponding
channels of the \mbox{$\Lambda$-cube} (correction based on field
stars, compensation of the offset between the calibration cube and
test interferograms of the rings from the calibration lamp before
and after observations of the
object ~\citep{Moiseev2002ifp}). Each channel  $I(x,y,z)$
thus stores correct wavelengths corresponding to each image pixel,
up to interpolation errors. The $I(x,y,z)\rightarrow
I(x,y,\lambda)$ conversion is performed by interpolating the
observed intensities at each point $(x,y)$ from the \mbox
{\mbox{$\Lambda$-cube}} scale to a uniform wavelength grid.

\begin{figure*}
 \includegraphics[scale=0.85]{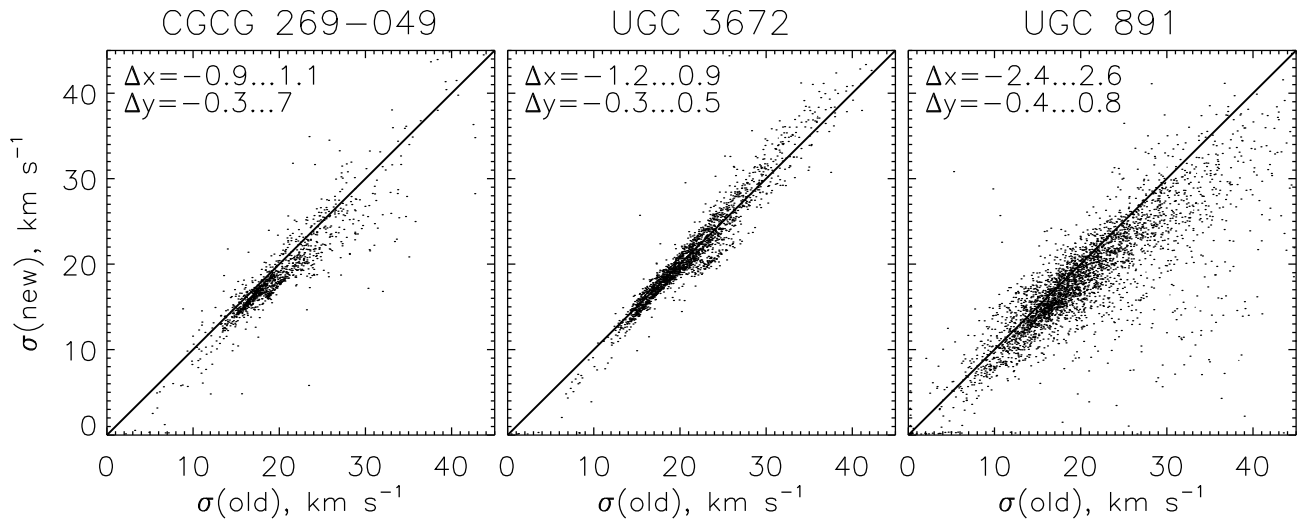}
\caption{Comparison of H$\alpha$  velocity dispersion measurements
for ionized gas in three dwarf galaxies (their names are given at
the top), made with IFP751. The wavelength scales along the
horizontal and vertical axes are based on the phase map
$\sigma({\rm old})$ and the  \mbox{$\Lambda$-cube} $\sigma({\rm
new})$ respectively. The dots show the measurements in each  data
cube pixel. The solid lines corresponds to the bisector
$\sigma({\rm old})=\sigma({\rm new})$. The numbers in the top left
corner indicate the range of mutual channel offsets in pixels
along the $x$ and $y$ axes relative to the mean position averaged
over all observations. One pixel corresponds to  $0\farcs7$.}
\label{fig6}
\end{figure*}

Figure~\ref{fig6} compares ionized gas velocity
dispersion measurements in three nearby dwarf galaxies
from~\citet{Moiseev2015} performed using the phase map
($\sigma({\rm old})$) and the \mbox{$\Lambda$-cube} method
($\sigma({\rm new})$). Shown are the cases where mutual channel
offsets exceeded one pixel for technical reasons. Usually
$\sigma({\rm new})<\sigma({\rm old})$, because in the case of a
nonzero offset the phase map method introduces extra scatter in
$\lambda$ resulting in a broadening of measured lines. However,
line profile distortions may be even more intricate,  because they
depend on the position of the target relative to the optical axis,
direction and magnitude of the offset, etc. That is why
$\sigma({\rm new})>\sigma({\rm old})$ in some areas. However, in
the cases of channel offsets no greater than~$1''$, the systematic
error of velocity dispersion measurements usually does not exceed
10~$\km$, i.e., the width of the instrumental profile \mbox{${\rm
FWHM}\approx2.35\sigma$}. Notice that measured line barycenter
positions are almost always free of systematic errors.

\section{Discussion}

\label{sec5}

Since 2011 the \mbox{$\Lambda$-cube} based calibration method has
been incorporated  into the IFPWID software
package~\citep{MoiseevEgorov2008}, written in IDL. It
avoids emission-line profile distortions arising in the cases of
individual channel offsets caused by technical problems with the
guiding system or spectrograph mechanisms. Although, as shown
above, the distortions lead mainly to the line profile broadening,
they prove to be critical in the tasks where the aim is not just
to determine the velocity field but also to perform a detailed
study of the line profile. A number of studies carried out on the
SAO~RAS 6-m telescope in recent years belong to this category,
e.g., the measurement of ionized gas velocity dispersion in dwarf
galaxies~\citep{Moiseev2015} or the study of young
stellar object outflows~\citep{Movsessian2015}.

The proposed wavelength scale calibration method is not
fundamentally new. \citet{BlandTully1989} pointed out in their classic
paper the need for individual phase correction of the channels in
the cases of ring center offsets. \citet{Mitchell2015} use a similar idea in their
analysis of observations made with the FPI on the 10-m Southern
African Large Telescope (SALT). In this case the wavelength
distribution in each channel is constructed based on the
preliminary measurement of the calibrating line (or the
measurements of the rings from night sky emission lines) and the
determination of several constants characterizing the appearance
of the fringe pattern \citep[see also][]{Gordon2000}.

At the same time, the use of the \mbox{$\Lambda$-cube} makes our
reduction programs easily adaptable to the analysis of the data
obtained with FPIs on other telescopes. In particular, we used this
technique to reduce the observations obtained with the scanning
FPI mounted on SALT~\citep{Brosch2015}. The fundamental
difference from observations performed at SAO~RAS is that because
of various technical limitations on the exposure duration, the
data cube cannot be  scanned over the entire overlap-free
$\Delta\lambda$ range but only in the region where the location of
an emission line is expected.

We hope that the parameters of the interferometers used with \mbox
{SCORPIO-2} and the modifications of the data reduction technique
described in this paper would be useful for colleagues interested
in the study of the kinematics of extended objects with the
scanning FPI on the SAO~RAS 6-m telescope.

\begin{acknowledgements}
This work was supported by grant MD-\linebreak 3623.2015.2 of the
President of the Russian Federation and by the non-profit Dynasty
Foundation. This paper is based on the observations carried out on
the 6-m telescope of the Special Astrophysical Observatory,
operated with the financial support of the Ministry of Education
and Science of the Russian Federation (agreement No.
14.619.21.0004, project~ID RFMEFI61914X0004). I am grateful to
IC~Optical Systems,~Ltd. and personally to Chris Pietraszewski for
sharing technical information and sustained interest in how their
interferometers are used in observations on the 6-m telescope.
\end{acknowledgements}


\end{document}